\documentclass[aps,prl,reprint,superscriptaddress]{revtex4-2}
\usepackage{amsmath,amssymb}
\usepackage{graphicx}
\usepackage{hyperref}
\usepackage{color}
\usepackage[normalem]{ulem}

\begin{document}
\raggedbottom

\newcommand{\martin}[1]{\textcolor{red}{#1}}

\title{From second to infinite order: Ehrenfest transitions in a one-dimensional hard-sphere crystal under compression}

\author{Sakineh Mizani}

\affiliation{Institute for Applied Physics, University of T\"ubingen, Auf der Morgenstelle 10, 72076 T\"ubingen, Germany}

\author{Szabolcs Varga}

\affiliation{Physics Department, Centre for Natural Sciences,
University of Pannonia, P.O. Box 158, Veszpr\'em H-8201, Hungary}

\author{Martin Oettel}

\affiliation{Institute for Applied Physics, University of T\"ubingen, Auf der Morgenstelle 10, 72076 T\"ubingen, Germany}

\author{P\'eter Gurin}

\affiliation{Physics Department, Centre for Natural Sciences,
University of Pannonia, P.O. Box 158, Veszpr\'em H-8201, Hungary}

\date{\today}

\begin{abstract}
Ehrenfest phase transitions, defined by the first discontinuity in the $n^{\mathrm{th}}$ derivative of the free energy, are rare and poorly understood for $n>2$. We present an exactly solvable one-dimensional model exhibiting a sequence of Ehrenfest transitions of arbitrary order $n\geq2$. These transitions occur without symmetry breaking and even the second-order transition lacks conventional critical behavior, although the density derivative of the correlation length diverges. Our results reveal a distinct class of noncritical phase transitions and help clarify the scope of the Ehrenfest classification.
\end{abstract}

\maketitle

Phase transitions are associated with singularities in the free energy. In an influential early paper, Ehrenfest \cite{Ehrenfest_1933} suggested a classification of first and second-order transitions through discontinuities of first and second derivatives of the free energy, respectively, which suggests a natural extension to  $n^{\text{th}}$-order transitions (as was later done in textbooks, e.g., by Pippard \cite{Pippard_1964}). The discovery of the roles of symmetry, topology, scaling and critical fluctuations in phase transitions has superseded Ehrenfest's classification and led to a classification scheme which basically distinguishes only between discontinuous (first order in the strict Ehrenfest sense) and continuous transitions \cite{Stanley_1971}. A nice historical overview of the subject can be found in Ref.~\cite{Jaeger_1998}. Although phenomenological theories such as Ginzburg--Landau theory can readily be constructed to exhibit Ehrenfest transitions of arbitrary order, exactly solvable models displaying such behavior remain scarce, and experimental evidence is even scarcer. Hence, genuine $n^{\text{th}}$-order transitions in the Ehrenfest sense ($n > 2$)  have remained elusive, with few exceptions: the condensation of the ideal Bose gas, the Gross-Witten-Wadia transition in the 2-dimensional $U(N)$ lattice gauge theory when $N\rightarrow\infty$ \cite{Gross-Witten_1980,Wadia_1980}, both of which are $n=3$, and the Ising model on the Bethe lattice , which shows both $3^{\text{rd}}$ and $4^{\text{th}}$-order transitions~\cite{MullerHartmann-Zittarz_PRL_1974,Stosic-Stosic-Fittipaldi_2009}.

In this Letter we present an exactly solvable 1D model which exhibits \textit{all} of the elusive Ehrenfest transitions. This is particularly interesting because well-known no-go theorems restrict the occurrence of phase transitions in 1D systems. The van Hove theorem\cite{van1950integrale,Cuesta-Sanchez_JStatPhys_2004} excludes thermodynamic phase transitions in a broad class of homogeneous 1D systems with short-range interactions, while the Mermin-Wagner theorem\cite{mermin1966absence} prohibits the spontaneous breaking of continuous symmetries under similar conditions. However, the assumptions underlying these theorems need not hold in all physically relevant realizations. For example, external periodic potentials can alter 1D systems and enable phase transitions. Such situations can be realized experimentally in carbon nanotube bundles and systems confined by optical lattices \cite{Tatarkova-Carruthers-Dholakia_PhysRevLett_2002,carraro2002existence,greiner2008optical,colodrero2008nanoparticle}, where spatial ordering can occur, with tuning of the lattice spacing readily achieved via external field parameters or optical trapping potentials. As the lattice constant decreases, the available volume for individual particle fluctuations shrinks, drastically amplifying steric constraints and the frequency of collisions between adjacent neighbors.

\begin{figure}[t]
\centering
\includegraphics[   width=0.95\columnwidth
    ]{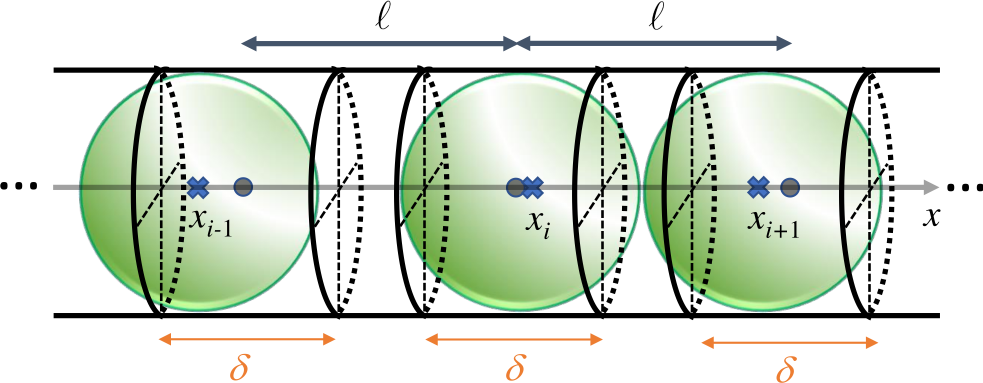}
\caption{Schematic of the 1D anchored hard-sphere solid. Hard spheres of diameter $\sigma$ move within intervals of length $\delta$ centered on lattice sites (grey dots, spacing $\ell$). Crosses mark particle positions, $x_{i}\in[-\delta/2,\delta/2]$. Particles cannot pass one another, preventing intersite hopping in this single-file system.}
\label{fig:fig1}
\end{figure}

Here we examine a model of a 1D solid consisting of hard spheres with diameter $\sigma$ confined by a periodic external potential (with lattice constant $\ell$) which consists of wells of width $\delta$ within which centers of particles can move freely, otherwise the external potential is infinite and thus the spheres cannot escape their wells, see Fig.~\ref{fig:fig1} {for a sketch}.
These wells model the lattice sites.
The number of the spheres, $N$, equals the number of the wells, so there is exactly one particle per site, since we assume $\delta<\sigma$. The hard spheres can interact with each other when $\ell<\delta+\sigma$. 

Below we determine the free energy per particle as a function of the particle number density $\rho=1/\ell$ and find a sequence of phase transitions in the strict Ehrenfest sense, starting with $n=2$ and going up to $n \to \infty$. The physical reason for the appearance of these transitions is that upon compression of the lattice, at a transition of order $n$, the character of configurations changes: the maximal number of spheres being possibly in contact in allowed configurations changes from $n-1$ to $n$. Thus, the Ehrenfest sequence of transitions is tied to the spread of correlations as the system is compressed.

This model of a 1D solid can be examined exactly using the transfer operator method in the canonical ensemble \cite{casey1969model,rojas2019universality,dauxois1995entropy}. Denoting the position of the center of the $i$th particle relative to the center of the $i$th well by $x_i$, the configurational part of the canonical partition function is given by:
\begin{equation}
Z_{N}=\int_{-\delta/2}^{\delta/2}\dots\int_{-\delta/2}^{\delta/2}\prod_{i=1}^{N-1}K(x_{i},x_{i+1})dx_{1}\dots dx_{N} \, ,
\end{equation}
where the Boltzmann factor $K(x,x')=\Theta(x'-x+d)$ is the step-function, ensuring that neighboring hard spheres do not overlap. Here, $d=\ell-\sigma$ is the average free space between neighbors. The thermodynamic properties of the system are determined by the solutions of the eigenvalue equation,
\begin{align}
  \int_{-\delta/2}^{\delta/2}K(x,x')\psi_k(x')dx' &= \lambda_k\psi_k(x),
  \label{eigenvalue_eq}
\end{align}
which gives the right hand side eigenfunctions, $\psi_k$, and the corresponding eigenvalues, $\lambda_k$. The integral operator defined by the kernel above---known as the transfer operator---is positive, compact, and irreducible~\cite{SM}, therefore, according to the Perron-Frobenius-Jentzsch theorem~\cite{Cuesta-Sanchez_JStatPhys_2004}, the dominant eigenvalue, $\lambda_{0}$, is non-degenerate and separated by a gap from the rest of the spectrum ($\lambda_{0}>|\lambda_{1}|\ge\dots$). Therefore, in the limit {$N\rightarrow\infty$}, $Z_{N}$ is dominated by the leading eigenvalue, $Z_{N}\sim\lambda_{0}^{N}$, yielding the free energy per particle $f=\beta F/N=-\ln\lambda_{0}$.

The local positional distribution function of a particle within its well, $\alpha(x)$, is given by the product of the r.h.s. and l.h.s. eigenfunctions corresponding to $\lambda_0$, which are different, because the kernel is not symmetric. However, due to the spatial inversion symmetry of the system, we have $K(-x',-x)=K(x,x')$, and based on this property we can write $\alpha(x)=\psi_0(-x)\psi_0(x)$,  normalized by $\int_{-\delta/2}^{\delta/2}\alpha(x)dx=1$. Furthermore, the positional correlation length $\xi$ follows directly from the spectral gap between the two leading eigenvalues:
\begin{equation} 
\xi^{-1} = \ln\left(\frac{\lambda_0}{|\lambda_1|}\right).
\label{eq:co}
\end{equation}

As we shall show, nonanalyticities in the eigenvalue problem can arise when the maximum number of consecutive contacting particles changes. Geometrically, for $n$ consecutive hard spheres to be in contact forming a chain, the distance between the first particle being at $+\delta/2$ and the $n^{\mathrm{th}}$ particle being at $(n-1)\ell-\delta/2$ must equal $(n-1)\sigma$, giving the threshold spacing $\ell=\sigma+\delta/(n-1)$, or equivalently $n=1+\delta/d$. Since this quantity is generally noninteger for an arbitrary $\ell=1/\rho$, we define $n(\rho)$ and $q(\rho)$ as the integer and fractional parts of $1+\delta/d$, respectively. Here $n(\rho)$ determines the maximum number of consecutive contacting particles, whereas $q(\rho)$ parametrizes the density interval over which $n$ remains constant. It is worth noting that, in a mathematical sense, $n$ refers to the smallest integer for which the $n^{\mathrm{th}}$ power of the integral operator defined by Eq.~(\ref{eigenvalue_eq}) is strongly positive{~\cite{SM}}.

When $n=1$ 
and $0\leq q<1$, corresponding to low densities, $\rho \leq 1/(\sigma+\delta)$, and a respective lattice constant $\ell>\delta+\sigma$, the particles can not interact with each other. Accordingly, $f=\text{const}$, since $K \equiv 1$, therefore, the only nonzero eigenvalue is $\lambda_{0} = \delta$, and the corresponding eigenfunction is also constant, showing that each sphere is uniformly distributed within its own well, since their neighbors do not influence their positions.

As the density increases and $n$ exceeds 1, the kernel vanishes whenever $x'<x-d$. Nevertheless, if $x<\Delta_1=-\frac{\delta}{2}+d$, then $K(x,x')=1$ for all $x'$. Consequently, Eq.~(\ref{eigenvalue_eq}) says that $\lambda_k\psi_k(x)=c_{k;0}$ is a constant for $x<\Delta_1$. In the regime $n=2$ ($0\leq q<1$), for $x\geq\Delta_1$ the integral in Eq.~(\ref{eigenvalue_eq}) can be decomposed into two contributions. For $x' <\Delta_1$, we have already established that $\psi_k(x')=c_{k;0}/\lambda_k$. For $x'\geq\Delta_1$, the corresponding contribution to the integral is constant, $c_{k;1}$, since $K=1$ throughout this domain. Therefore $\lambda_k\psi_k(x)=\int_{x-d}^{\Delta_1}\frac{c_{k;0}}{\lambda_k}  {\,dx'} +c_{k;1}=\frac{c_{k;0}}{\lambda_k}(\Delta_2-x)+c_{k;1}$, where  $\Delta_2=-\frac{\delta}{2}+2d$. Furthermore, for $\lambda \neq 0$, $\psi(x)$ must be a continuous function of $x$, since the boundary $x'=x-d$, which separates the $K=0$ and $K=1$ domains, is a continuous function of $x$.
Matching the solutions obtained on the two intervals at $x=\Delta_1$ determines $c_{k;1}$, yielding
\begin{equation}
  \psi_{k}(x)= 
  \begin{cases}
    \frac{c_{k;0}}{\lambda_k},    &\text{if } -\delta/2\le x<\Delta_{1},\\
    c_{k;0}\left(\frac{\Delta_{2}-x}{\lambda_k^2}
    +\frac{\lambda_{k}-d}{\lambda_k^2}\right) ,
          &\text{if }   \Delta_{1}\le x\le\delta/2,
  \end{cases}
\end{equation}
where $c_{k;0}$ serves as a normalization constant. Inserting $\psi_{k}(x)$ back into Eq.\ref{eigenvalue_eq} results in a quadratic characteristic equation for the eigenvalue:
\begin{equation}
  \left(\frac{\lambda_k}{d}\right)^2-(1+q)\frac{\lambda_k}{d}+\frac{q^2}{2}=0,
  \label{characteristic_eq-n=2}
\end{equation}
where we used that $q=\delta/d-1$ in the studied density regime corresponding to $n=2$.
The largest root of Eq.~(\ref{characteristic_eq-n=2}) determines the free energy per particle.

\begin{figure}[t]
\centering
\includegraphics[   width=\columnwidth,
    ]{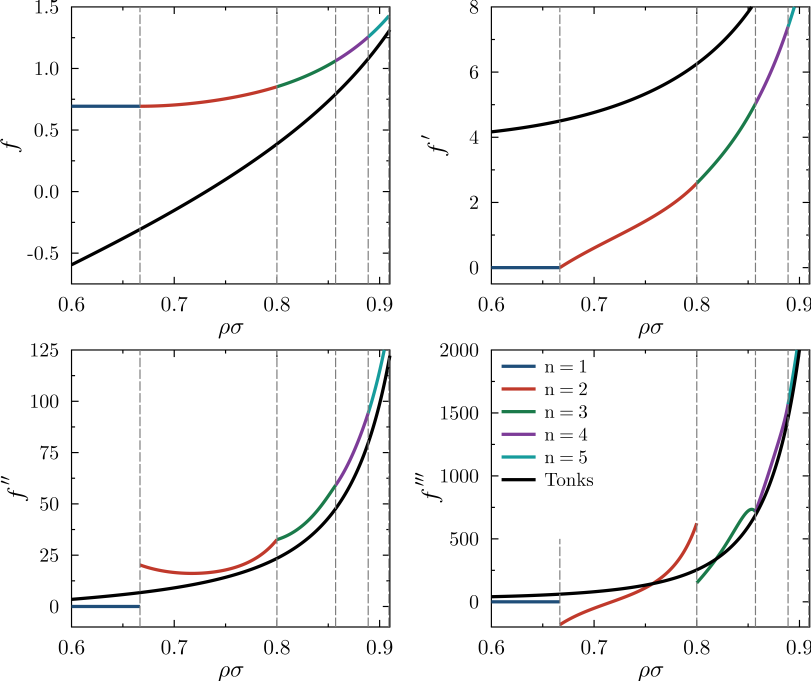}
\caption{Thermodynamic properties upon compression. $f$ and its first three reduced density derivatives ($f', f'', f'''$) versus reduced density $\rho\sigma$ for $\delta=\sigma/2$. Colored curves denote successive density regimes {specified by $n$} separated by vertical dashed lines at $\rho_{n}$. Black curves show the Tonks gas result, $f_{\text{Tonks}}=\ln\rho{\sigma}-1-\ln(1-\rho\sigma)$, for comparison.}
\label{fig:fig2}
\end{figure}

The generalization of the above results to higher densities is straightforward{~\cite{SM}}. The solutions of Eq.~(\ref{eigenvalue_eq}) for arbitrary $n$ and $q$ can be characterized as follows. All nonzero eigenvalues are obtained as roots of
\begin{equation}
  \sum_{i=0}^{n} \frac{(-1)^{n-i}(i+q)^{n-i}}{(n-i)!}
  \left(\frac{\lambda_k}{d}\right)^{i}=0,
  \label{characteristic_eq}
\end{equation}
while the corresponding eigenfunctions are piecewise polynomial functions.
For each value of $n$, Eq.~(\ref{characteristic_eq}) defines a distinct set of solutions $\tilde\lambda_k^{(n)}(q)$ for any $q\in\mathbb{R}$ (and not only for $0\leq q<1$). According to a mathematical theorem, a simple root of a polynomial is an analytic function of its coefficients{, therefore} any simple root of Eq.~(\ref{characteristic_eq}) is an analytic function of $q$. Consequently, $\tilde\lambda^{(n)}_0(q)$ is also analytic with respect to $q$, given that it is a non-degenerate eigenvalue (i.e., a simple root){~\cite{SM}}. Within a density interval characterized by fixed $n$, variations in the density are entirely encoded in $q$. Since $q$ is an invertible analytic function of $\rho$ on such an interval, analyticity with respect to the density is equivalent to analyticity with respect to $q$. Therefore the free energy remains analytic throughout every density interval where $n$ is fixed. The eigenvalues $\lambda_k$ are obtained by piecing together the branches $\tilde\lambda_k^{(n)}(q)$ corresponding to successive values of $n$. 
These branches are valid on different density intervals and {join} at 
\begin{equation}
 \rho_n=\frac{n-1}{(n-1)\sigma+\delta}\;,
\end{equation}
corresponding to $q=0$ and $q=1$ at the boundaries of adjacent intervals characterized by $n$ and $n-1$, respectively. \textit{At these densities, the branches connect non-analytically.} These findings are illustrated in Fig.~\ref{fig:fig2} for $n$ up to $5$. Specifically, discontinuities appear in $f''$ and $f'''$ at $\rho_2$ and $\rho_3$, respectively, while kinks appear in $f'$,  $f''$  and $f'''$ at $\rho_2$, $\rho_3$ and $\rho_4$, respectively. By contrast, {there is }no discontinuity or kink at $\rho_5$ in the shown derivatives of $f$. These results demonstrate the general result that the $n^{\text{th}}$ derivative of the free energy with respect to the density is discontinuous at $\rho_n$, while all lower-order derivatives remain continuous, marking a $n^{\text{th}}$-order phase transition in the Ehrenfest sense. Thus, compression drives an infinite series of phase transitions of sequentially higher order $n\ge2$, with all orders realized except for $n=1$. As $n$ increases, the transition densities $\rho_{n}$ accumulate toward the close-packing limit $\rho_{\mathrm{cp}}=1/\sigma$, so that neighbouring transition densities get closer to each other.

 We explicity illustrate the general result in the simplest case $n=2$. From Eqs.~(\ref{characteristic_eq-n=2}) or (\ref{characteristic_eq}), it follows that, for $(n,q)=(2,0)$, $\lambda_0$ is the largest root of $\frac{\lambda_k}{d}(\frac{\lambda_k}{d}-1)=0$ whereas for $(n,q)=(1,1)$, $\lambda_0=\delta$ is the root of $\frac{\lambda_k}{d}-1=0$. Hence, $\lambda_0$ remains continuous across the transition from $(n,q)=(1,1)$ to $(2,0)$. Consequently, the free energy $f$ is continuous at $\rho_2=\frac{1}{\sigma+\delta}$.
It can also be seen that the subleading eigenvalue $\lambda_1$ emerges from zero at this point and increases for larger densities. Furthermore, throughout the $n=1$ regime, all derivatives of $\lambda_0$ (and therefore of $f$) vanish. By contrast, Eq.~(\ref{characteristic_eq-n=2}) yields $\lambda_0=\frac{d}{2}(1+q+\sqrt{1+2q-q^2})$. Expanding around $q=0$, one finds that the first derivative of $f=-\ln\lambda_0$ vanishes at the transition point, whereas the second derivative remains finite and nonzero, $f''=(\sigma+\delta)^{4}/\delta^{2}$, marking a second-order phase transition in the Ehrenfest sense.

\begin{figure}[t]
  \centering
    \includegraphics[width=\columnwidth
    ]{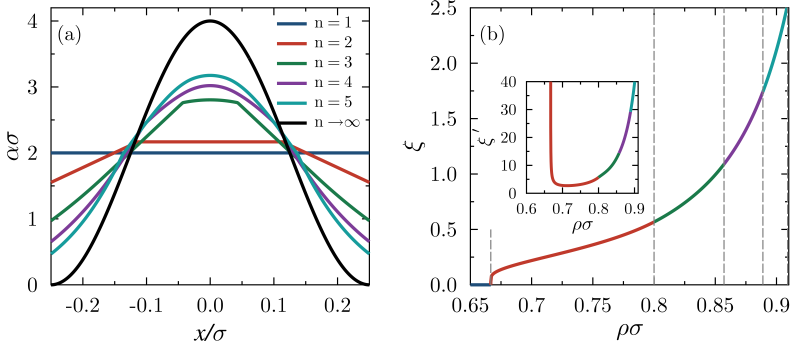}
  \caption{Local order and correlation length. (a) Local positional probability distributions
    $\alpha\sigma$ at interval midpoint densities
    $\bar{\rho}_{n}=(\rho_{n}+\rho_{n+1})/2$ for
    $\delta{/\sigma}=0.5$. {(b)} Positional correlation
    length $\xi$ (in units of particle number)
    versus density for $\delta{/\sigma}=0.5$. Inset: Density
    derivative $\xi'=d\xi/d(\rho\sigma)$ versus $\rho\sigma$}.
  \label{fig:fig3-4}
\end{figure}

Fig.~\ref{fig:fig3-4}{(a)}
illustrates the evolution of $\alpha(x)$ with increasing density. In the non-interacting regime ($n=1$), $\alpha(x)$ is strictly uniform across the interval $[-\delta/2,\delta/2]$. As the system is compressed into the $n=2$ regime ($\rho>\rho_{2}$), hard-core collisions set in, giving rise to a pair of kink points in $\alpha(x)$ at $x=\pm\Delta_1$. Each subsequent threshold density $\rho_{n}$ introduces two additional singular points, leading to a $(2n-1)$-segment polynomial profile{~\cite{SM}}.  In the close-packing limit $n\rightarrow\infty$ ($\rho\rightarrow1/\sigma$), the distribution converges to the continuous cosine-squared form:
\begin{equation}
\alpha(x)=\frac{2}{\delta}\cos^{2}\left(\frac{\pi x}{\delta}\right).
\end{equation}
Crucially, the distribution remains finite and smooth rather than collapsing into a Dirac delta function, demonstrating that positional fluctuations persist even at close packing ($\rho\rightarrow 1{/\sigma}$) via concerted interparticle motion.

The onset of hard body collisions also affects the distance over which positional fluctuations propagate across lattice sites. Fig.~\ref{fig:fig3-4}{(b)} displays the positional correlation length $\xi$ (see Eq.~\eqref{eq:co})  alongside its density derivative $\xi' = d\xi/d\rho$ (inset). In the non-interacting regime ($n = 1$), displacements in neighboring cells are completely independent, corresponding to $\xi = 0$. Upon approaching the second-order transition threshold from above ($\rho = \rho_2 + \epsilon$ with $\epsilon \to 0^+$), the spectral gap behavior yields the asymptotic forms:
\begin{equation}
\xi \sim \frac{1}{2|\ln \epsilon|}, \quad \text{and} \quad \frac{d\xi}{d\rho} \sim \frac{1}{2\epsilon (\ln \epsilon)^2}.
\label{eq:xi_asymptotics}
\end{equation}
These equations show that $\xi \to 0$ for $\epsilon \to 0^+$, confirming that no diverging critical fluctuations accompany the second-order transition. Conversely, $d\xi/d\rho \to \infty$ for $\epsilon \to 0^+$, signaling the singular onset of spatial correlations across the 1D crystal. Beyond this onset, the rapid initial growth of $\xi$ quickly weakens, manifesting as a minimum in $\xi'$ within the second density region (right panel of Fig.~\ref{fig:fig3-4}, inset).
Upon passing into the 3rd, 4th, and higher density regions, positional correlations grow systematically. This strengthening is directly linked to steric confinement: the maximum number of particles that can be in simultaneous contact is identical with $n$, which determines the density region and the order of phase transition as well. Consequently, the correlation length remains strictly bounded by $n$ throughout the $n^{\text{th}}$ density region ($\xi < n$). As close packing is approached ($\rho\rightarrow 1/\sigma$), these expanding multi-particle contact constraints drive $\xi$ to grow monotonically, meaning that an ever-larger number of neighboring particles must move in concert to sustain positional fluctuations within the crystal.

The origin of this unusual phase sequence---characterized by weakening singularities without continuous symmetry breaking---lies in the progressive formation of clusters under compression. Rather than undergoing a global structural transformation, the crystal evolves through an infinite sequence of
Ehrenfest transitions purely as a function of $n$ being identical with the maximum cluster size. The correlation length $\xi$ does not diverge at the transition boundaries; instead, $\xi$ grows continuously across higher-order transition densities, while higher-order generalized correlation lengths
$\xi_{0,n-1} := 1/\ln(\lambda_0/|\lambda_{n-1}|)$
 inherit derivative singularities corresponding to the opening of higher-order phase transitions.

It is interesting to note that the singularities appear to survive in the corresponding grand canonical (gc) system where the potential wells are placed in a bath with chemical potential $\mu$. The singularities appear for $\mu=\text{const.}$ and decreasing well spacing $\ell$.  (The analyzed system above corresponds to the gc system with $\mu\to\infty$, leading
 to occupation one in each well.) The grand canonical system is exactly treatable through classical density functional theory and the minimization of its grand free energy functional using Percus’ exact free energy functional \cite{percus1976equilibrium}. The analytic solutions for the regimes $n=1$ and $n=2$ exhibit the jump in the second derivative of $F/N$ as before~\cite{SM}. This is indeed surprising since for low $\mu$ the wells are hardly populated and mostly vacant. In numerical solutions of the minimizing Euler-Lagrange equation we see also evidence of higher-order singularities and hence we  speculate that also all higher transitions will be found in the gc system.

Rather than signaling conventional thermodynamic phase transitions, the singularities identified here reflect a sequence of geometrical phase transitions driven by steric constraints. Each density interval $[\rho_{n-1},\rho_n]$ hosts a uniquely stable phase characterized by a maximal contact-chain length $n$. Because the free energy remains strictly analytic within each interval, internal transitions and phase-coexistence gaps are prohibited; transitions occur exclusively at the boundaries $\rho_n$ where different branches join non-analytically. These transitions lack critical behavior or diverging correlation lengths. Instead, the singularities in the free-energy derivatives act as quantitative measures of structural differences between neighboring phases. This demonstrates that the Ehrenfest classification serves as a powerful diagnostic tool for identifying structural changes in geometrically constrained systems. More broadly, the present framework may admit a natural generalization to higher dimensions: in two- and three-dimensional crystal lattices, tuning the lattice spacing at fixed positional fluctuations could generate an infinite hierarchy of distinct phases with controllable multiparticle cluster sizes, offering a general framework for quantifying structural differences through Ehrenfest derivatives.

Regarding the connection to other 1D systems, an infinite number of transitions (without, however, the Ehrenfest sequence) are known in energy-driven (thermal) models, involving specially designed many-body interactions \cite{saryal2018multiple,klamser2022sequence} or in devil's staircase phenomena in thermal ground states of systems governed by incommensurability \cite{fisher1980infinitely,Aubry1983,bak1982one}. In contrast, the present model is purely entropy-driven and built on the simplest hard body interaction.

From an experimental perspective, the model system studied here can be realized by confining colloidal particles within narrow nanochannels, where single-file dynamics and nearly hard-body exclusions occur \cite{wei2000single,lin2005random}. Similar colloidal setups have been used to study the equation of state and spatial fluctuations of the 1D Tonks gas \cite{lin2009quasi}. Furthermore, using optical tweezers or periodic laser-trapping arrays, an anchored 1D colloidal solid subject to tunable local potentials can be engineered \cite{Tatarkova-Carruthers-Dholakia_PhysRevLett_2002} and tracked in real time via video microscopy \cite{greiner2008optical}.

As an outlook, the theoretical framework developed here has broad relevance to real-world one-dimensional and quasi-1D finite-length chains, such as double-stranded DNA molecules. In DNA, complementary base pairs are joined by hydrogen bonds, and thermal or force-induced denaturation—the strand-splitting transition—can be effectively described using the transfer operator method \cite{kittel1969phase,dauxois1993entropy}. By decorating our hard-sphere solid model with short-range hydrogen-bonding sites, this approach can be naturally extended to study how positional fluctuations, spatial confinement, and steric hard-body exclusions directly influence denaturation thermodynamics in finite biomolecular chains.
\section*{Acknowledgements}
S. M. acknowledges the financial support of Humboldt Foundation. P. G. and S. V. gratefully acknowledge the financial support of the National Research, Development, and Innovation Office - NKFIH K137720.

\bibliography{references}

\end{document}


\raggedbottom

\title{Supplemental Material\\ 
An infinite sequence of higher order Ehrenfest transitions in a one-dimensional crystal}

\author{P\'eter Gurin}

\affiliation{Physics Department, Centre for Natural Sciences,
University of Pannonia, P.O. Box 158, Veszpr\'em H-8201, Hungary}

\author{Szabolcs Varga}

\affiliation{Physics Department, Centre for Natural Sciences,
University of Pannonia, P.O. Box 158, Veszpr\'em H-8201, Hungary}

\author{Martin Oettel}

\affiliation{Institute for Applied Physics, University of T\"ubingen, Auf der Morgenstelle 10, 72076 T\"ubingen, Germany}
\author{Sakineh Mizani}

\affiliation{Institute for Applied Physics, University of T\"ubingen, Auf der Morgenstelle 10, 72076 T\"ubingen, Germany}

\date{\today}

\maketitle

\section{Positivity, compactness, irreducibility and analyticity of the transfer operator}

In this Section we prove that the integral operator defined in the main text, namely
\begin{align}
  (\hat{K}\psi)(x) = \int_{-\delta/2}^{\delta/2}K(x,x')\psi(x')dx'
  \label{eigenvalue_eq}
\end{align}
where
\begin{align}
    K(x,x')
  = \Theta(x'-x+d)
  = \begin{cases}
      0 \quad\text{if }x'\leq x-d\\
      1 \quad\text{if }x'> x-d\\
    \end{cases}
  \label{K}
\end{align}
is a positive, compact, and irreducible operator for any $d>0$.

An operator is called positive, if for any $\psi(x)\geq0$ it follows that $(\hat{K}\psi)(x)\geq0$, in other words, it maps a non-negative function to a non-negative function. It immediately follows, since the kernel is non-negative.

Furthermore, $\int_{-\delta/2}^{\delta/2} dx \int_{-\delta/2}^{\delta/2} dx' |K(x,x')|^2<\infty$, so $\hat K$ is a Hilbert--Schmidt operator and therefore compact.

To establish irreducibility, we first determine the support of the kernels of the powers of $\hat K$. We claim that, for every integer $m\geq1$,
\begin{equation}
  K^m(x,x')>0
  \qquad\Longleftrightarrow\qquad
  x'>x-md.
  \label{lemma}
\end{equation}
The proof is by induction on $m$. For $m=1$, the statement follows immediately from the definition of the kernel. Assume that for some $m\geq1$ the equivalence stated by Eq.~(\ref{lemma}) is true.
Using the composition rule of integral operators,
\begin{equation}
K^m(x,x')
=
\int_{-\delta/2}^{\delta/2} K^{m-1}(x,x'')K(x'',x')\,dx'' ,
\end{equation}
the integrand is positive precisely when $x''>x-(m-1)d$ and $x'>x''-d$. Equivalently, $x-(m-1)d<x''<x'+d$. Therefore $K^m(x,x')$ is positive if and only if the interval $\bigl(x-(m-1)d,\;x'+d\bigr)$ has positive length. Its length is $x'+d-\bigl(x-(m-1)d\bigr)=x'-x+md$. Hence $K^m(x,x')>0\iff x'-x+md>0$, which is equivalent with Eq.~(\ref{lemma}). This completes the induction.

Choose the integer $n=\left\lfloor\frac{\delta}{d}\right\rfloor+1$, which guarantees that $nd>\delta$. Since $-\frac{\delta}{2}\leq x,x'\leq\frac{\delta}{2}$, we have $x'-x\geq -\delta$. Therefore $x'-x+nd>-\delta+\delta=0$, and thus $x'>x-nd$. Consequently, $K^n(x,x')>0$ for all $-\frac{\delta}{2}\leq x,x'\leq\frac{\delta}{2}$.

Now let $\psi\geq 0$ be nonzero. With this $\psi$ we have $(\hat K^n\psi)(x)=\int_{-\delta/2}^{\delta/2} K^n(x,x')\psi(x')\,dx'>0$ for all $-\frac{\delta}{2}\leq x\leq\frac{\delta}{2}$, which means that $\hat K^n$ is strongly positive. Since a positive power of $\hat K$ is strongly positive (i.e. the operator $\hat K$ is primitive), therefore $\hat K$ is irreducible.

Since $\hat K$ is positive, compact, and irreducible, the Perron--Frobenius--Jentzsch theorem applies. Therefore, the dominant eigenvalue of the transfer operator cannot become degenerate. As a consequence, it does not cross with another eigenvalue, meaning that the correlation length never diverges, i.e., $\xi$ 
stays finite at every transition. However, the kernel is not an analytic function of the density. Consequently, the standard analyticity results for analytic families of linear operators do not apply, (i.e., a condition of Theorem 4 of Ref.~\cite{Cuesta-Sanchez_JStatPhys_2004} does not hold),
and nonanalytic behavior of the dominant eigenvalue is not excluded. This provides the mathematical explanation for the possibility of the singular behavior of the free energy, not the violation of the PFJ theorem.

\section{Exact solution of the transfer operator eigenvalue problem for general $n$}

In this section, we provide the mathematical details of the exact solution of the eigenvalue equation presented in Eq.~(2) of the main text.

As in the main text,  introducing the integer index $n=\lfloor\delta/d\rfloor+1$, and $0\leq q= \{\delta/d\} <1$, we separate the integer part and the fractional part of $\delta/d$, i.e. $\delta/d=n-1+q$. (Here $\lfloor\rfloor$ and $\{\}$ denote the integer- and fractional part, respectively, and $d=\ell-\sigma$.) Thus, for given values of $\delta$ and $\sigma$, $n$ and $q$ together determine the density (or the lattice constant) as $\rho=1/\ell=\frac{n-1+q}{(n-1+q)\sigma+\delta}$.

We shall see that the solutions of the eigenvalue equation split into distinct branches which, for a fixed value of $n$, depend analytically on $q$. However, when $n$ changes by one (and hence $q$ changes accordingly), these branches do not join analytically. To distinguish these branch solutions from the general solutions of the eigenvalue equation denoted by $\psi_k(x)$ and $\lambda_k$, we denote by $\tilde\psi^{(n)}_k(x)$ and $\tilde\lambda^{(n)}_k$ the eigenfunctions and eigenvalues associated with a given branch. Certainly, both of them depend on the density, which dependence can be taken into account as a dependence on $q$, as $n$ is fixed.

For a given $n$, using the specific form of the kernel defined in Eq.~(\ref{K}), the eigenvalue equation can be written as
\begin{equation}     
     \int_{\max\{-\delta/2,x-d\}}^{\delta/2} dx'\, \tilde\psi^{(n)}_k(x')     
  =  \tilde\lambda^{(n)}_k \tilde\psi^{(n)}_k(x) 
  .
  \label{eig}
\end{equation}

\begin{figure}
\resizebox{0.48\textwidth}{!}{\includegraphics{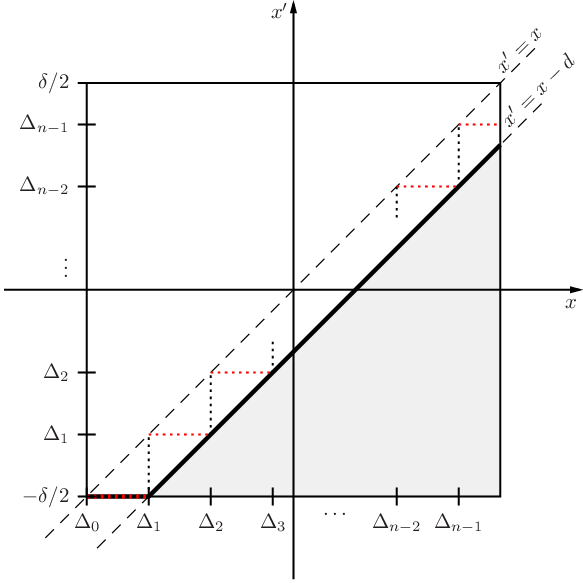}}
\caption{Schematic representation of the integration domain of the kernel $K(x,x')$. The kernel is defined on the domain $[-\frac{\delta}{2},\frac{\delta}{2}]\times[-\frac{\delta}{2},\frac{\delta}{2}]$. The shaded triangular region indicates the area where $K(x,x')=0$, while  $K(x,x')=1$ in the unshaded region. Thus, in the eigenvalue equation represented by Eq.~(\ref{eig}), the lower limit of integration with respect to $x'$ is marked by the thick line. On the domains located above the red dashed lines, the resulting integral is independent of $x$.
\label{fig:domains_of_eig._integral_eq.}}
\end{figure}

Introducing the notation $\Delta_i=-\delta/2+id$, it follows immediately from Fig.~\ref{fig:domains_of_eig._integral_eq.} and the eigenvalue equation \ref{eig} that $\tilde\lambda^{(n)}_k \tilde\psi^{(n)}_k(x)=c^{(n)}_{k;0}$ is constant for $\Delta_0\leq x <\Delta_1$, where
$c^{(n)}_{k;0}=\int_{-\delta/2}^{\delta/2} dx'\, \tilde\psi^{(n)}_k(x')$.
The figure also shows that, for $\Delta_i\leq x <\Delta_{i+1}$, the integral appearing in the eigenvalue equation \ref{eig} can be split into two contributions. Over the interval $[x-d,\Delta_i]$, the value of the integral depends on $x$ through the lower integration limit, whereas over $[\Delta_i,\delta/2]$ the integral is a constant independent of $x$; we denote this constant by $c^{(n)}_{k;i}$. Consequently, on successive intervals $[\Delta_i,\Delta_{i+1}]$ of length $d$, the segments of the eigenfunction satisfy the recurrence relation
\begin{equation}     
     \tilde\lambda^{(n)}_k P^{(n)}_{k;i}(x)    
  =  \int_{x-d}^{\Delta_i} dx'\, P^{(n)}_{k;i-1}(x') + c^{(n)}_{k;i}
  ,
  \quad i=1,\dots,n-1
  \label{rec}
\end{equation}
where we introduced the following notation for the segments of the eigenfunction,
\begin{equation}
      P^{(n)}_{k;i}(x)
  \equiv  \tilde\psi^{(n)}_k(x) \quad \text{if} \quad x\in[\Delta_i,\Delta_{i+1}]
 \label{P<-->psi}
\end{equation}

If we seek solutions with $\tilde\lambda^{(n)}_k \neq 0$, then, since $\tilde\psi^{(n)}_k(x)= P^{(n)}_{k;0}(x)=c^{(n)}_{k;0}/\tilde\lambda^{(n)}_k$ on the first interval, the above recursion implies that $P^{(n)}_{k;i}$ is a polynomial of degree $i$. Therefore, we seek polynomials that represent the eigenfunctions corresponding to the nonzero eigenvalues, segment by segment, in the following form:
\begin{equation}     
     P^{(n)}_{k;i}(x)    
  =  \sum_{j=0}^i a^{(n)}_{k;i,j} (\Delta_{i+1}-x)^j
  ,
  \label{P}
\end{equation}
%
Substituting Eq.~(\ref{P}) into Eq.~(\ref{rec}) yields the recurrence relations
\begin{align}
     a^{(n)}_{k;i,0} = \frac{c^{(n)}_{k;i}}{\tilde\lambda^{(n)}_{k}}, \qquad 
     a^{(n)}_{k;i,j} = \frac{a^{(n)}_{k;i-1,j-1}}{j\cdot\tilde\lambda^{(n)}_{k}}
        \quad \text{if } j>0
  \label{rec_a}
\end{align}
for the coefficients defining the polynomials $P_i$. Repeated application of the second relation in Eq.~(\ref{rec_a}) gives
\begin{equation}
    a^{(n)}_{k;i,j}
  = \frac{a^{(n)}_{k;i-j,0}}{j!(\tilde\lambda^{(n)}_{k})^j} \quad \text{if } j>0
  .
  \label{a^{(n)}_{k;i,j}}
\end{equation}

Since the boundary separating the regions $K(x,x')=0$ and $K(x,x')=1$ in the kernel, namely $x'=x-d$, is a continuous function of $x$, and since the eigenfunction is bounded (being piecewise polynomial), it follows from Eq.~(\ref{eig}) that any eigenfunction corresponding to a nonzero eigenvalue must itself be continuous. Continuity of $\tilde\psi^{(n)}_k$ at $x=\Delta_i$ implies
\begin{equation}
     P^{(n)}_{k;i-1}(\Delta_i)
  =
     P^{(n)}_{k;i}(\Delta_i)
  \, .
  \label{continuity}
\end{equation}
Using this condition together with the form (\ref{P}) of the polynomial we can eliminate $c^{(n)}_{k;i}$ from the recurrence relation (\ref{rec_a}),
\begin{equation}
    a^{(n)}_{k;i,0}
  = a^{(n)}_{k;i-1,0}
   -\sum_{j=1}^i a^{(n)}_{k;i-1,j-1} \frac{d^j}{j\cdot\tilde\lambda^{(n)}_{k}}
  .
  \label{without_c}
\end{equation}
Using Eq.~(\ref{a^{(n)}_{k;i,j}}) in Eq.~(\ref{without_c}) yields a linear recurrence relation for $a^{(n)}_{k;i,0}$,
  \begin{equation}
    a^{(n)}_{k;i,0}
  = a^{(n)}_{k;i-1,0}
   -\sum_{j=1}^i a^{(n)}_{k;i-j,0} \;\frac{1}{j!}\left(\frac{d}{\tilde\lambda^{(n)}_{k}}\right)^j
  ,
  \label{rec_for_a^{(n)}_{k;i,0}}
\end{equation}
which can be solved in closed form by means of the standard generating-function method, and using again Eq.~(\ref{a^{(n)}_{k;i,j}}) one obtains 
\begin{equation}     
     a^{(n)}_{k;i,j}    
  =  a^{(n)}_{k;0,0} \frac{1}{j!(\tilde\lambda^{(n)}_{k})^j} \sum_{l=0}^{i-j} \frac{(l+j-i-1)^l}{l!} \left(\frac{d}{\tilde\lambda^{(n)}_{k}}\right)^l
  ,
  \label{a}
\end{equation}
which, together with Eqs.~(\ref{P<-->psi}-\ref{P}) gives the exact eigenfunction, $\tilde\psi^{(n)}_k(x)$, in the density range $\rho_n\leq\rho<\rho_{n+1}$, where $\rho_n=\frac{n-1}{(n-1)\sigma+\delta}$. In Eq.~(\ref{a}) $a^{(n)}_{k;0,0}$ serves as a normalization constant for $\tilde\psi^{(n)}_k$.

Up to this point, the eigenvalue equation has been used through Eq.~(\ref{rec})---together with the continuity condition (\ref{continuity})---to recursively construct successive segments of the eigenfunction. In the final step, once the eigenfunction has been defined over the entire interval $[-\delta/2,\delta/2]$, Eq.~(\ref{rec}) becomes a genuine eigenvalue equation, since the constant term $c^{(n)}_{k;n-1}$ appearing on its right-hand side can now be computed from the known polynomial $P^{(n)}_{k;n-1}$. The eigenvalue is therefore determined from Eq.~(\ref{rec}) evaluated at $x=\Delta_{n-1}$:
\begin{align}     
     \tilde\lambda^{(n)}_k 
  =
     \int_{\Delta_{n-2}}^{\Delta_{n-1}} dx'\, P^{(n)}_{k;n-2}(x')
   + \int_{\Delta_{n-1}}^{\delta/2} dx'\, P^{(n)}_{k;n-1}(x')
   ,
   \label{last_rec}
\end{align}
Using the forms of the polynomials and coefficients given by Eqs.~(\ref{P}) and (\ref{a}), this equation can be reduced to
\begin{equation}     
     \sum_{i=0}^n \frac{(-1)^{n-i}(i+q)^{n-i}}{(n-i)!} \left(\frac{\tilde\lambda^{(n)}_k}{d}\right)^i   
  =  0
  .
  \label{lambda}
\end{equation}
The eigenvalues in the density range $\rho_n\leq\rho<\rho_{n+1}$ are given by the roots of the above $n^{\text{th}}$-degree polynomial.

Once these roots are known, the individual segments of the corresponding eigenfunctions follow from Eqs.~(\ref{a}) via (\ref{P<-->psi}) and (\ref{P}). We have seen that $\tilde\psi^{(n)}_k$ is a piecewise polynomial function with non-analytic points at $\Delta_i$, where $i=1,\dots,n-1$. Therefore $\alpha(x)=\psi_0(-x)\psi_0(x)$ has $2(n-1)$ non-analytic points at $\pm\Delta_i$, and between these kink points, $\alpha$ is a polynomial function of $x$.

\subsection{Eigenfunctions with zero eigenvalue}

The piecewise polynomial form of Eq.~(\ref{P}) follows from the recursion Eq.~(\ref{rec}) when $\tilde\lambda^{(n)}_k \neq 0$. Now we show that there are further eigenfunctions corresponding to zero eigenvalue, we index them by $\bar k$ simply to distinguish them from the previously presented solutions. In this case the left hand side of Eq.~(\ref{rec}) is zero, and $c^{(n)}_{\bar k;i}$ is an $x$ independent constant. However, the integral can not be independent of $x$ only in the case when the integrand is zero, i.e. $P^{(n)}_{\bar k;i-1}(x')=0$ for any $i=1,\dots,n-1$. Therefore only $P^{(n)}_{\bar k;n-1}(x')$ can differ from zero, because the recurrence formula for this case is (see Eq.~(\ref{last_rec}) with $\tilde\lambda^{(n)}_{\bar k}=0$ and $P^{(n)}_{\bar k;n-2}=0$) the following:
\begin{align}     
     0 = \int_{\Delta_{n-1}}^{\delta/2} dx'\, P^{(n)}_{\bar k;n-1}(x').
\end{align}
Therefore, when the eigenvalue is zero, only the last segment of the function in the domain $[\Delta_{n-1},\delta/2]$ can be different from zero, but its integral must be zero. Therefore a possible basis in the kernel of the operator $\hat K$ is the following:
\begin{align}
    \psi^{(n)}_{\bar k}(x)
 &= \begin{cases}
      0 \quad &\text{if } x<\Delta_{n-1}\\
      e^{i \bar k \frac{2\pi}{\delta/2-\Delta_{n-1}}(x-\Delta_{n-1})}  \quad &\text{if } x\geq\Delta_{n-1}
    \end{cases},
\end{align}
where $\bar k\in\mathbb{N}\setminus\{0\}$. We note that these eigenfunctions do not have to be continuous.

\section{Analytic Properties of the Free Energy}

Based on the results established above, we now provide the mathematical proof of the central claim of this paper: the $n^{\text{th}}$ derivative of the free energy is discontinuous at $\rho_n$, while all derivatives of lower order remain continuous.

Within a given density regime characterized by $n$, variations in the density can be mapped onto variations in $q$. Recall that, within such a density interval, the relation between $q$ and $\rho$ is invertible and analytic,
\begin{equation}
    q(\rho) = \frac{\delta}{1/\rho-\sigma}-n+1,
    \qquad
    \rho(q) = \frac{n-1+q}{(n-1+q)\sigma+\delta}.
\end{equation}
Therefore, the eigenvalues $\lambda_k$, and hence the free energy $f=-\ln\lambda_0$, are analytic functions of $\rho$ if and only if $\tilde\lambda_k^{(n)}$ is an analytic function of $q$, since by definition $\lambda_k\equiv\tilde\lambda_k^{(n)}$ throughout the density interval associated with $n$. Using Eq.~(\ref{lambda}), we may write
\begin{equation}
     \tilde\lambda^{(n)}_k
  =  d\cdot y^{(n)}_k ,
 \label{lambda<-->x}
\end{equation}
where $y^{(n)}_k(q)$ is implicitly defined as a root of the polynomial
\begin{equation}
    Q_n(y,q)
  = \sum_{i=0}^n (-1)^{n-i} \frac{(i+q)^{n-i}}{(n-i)!}\,y^i
  ,
 \label{Q}
\end{equation}
that is, $Q_n(y_k^{(n)}(q),q)=0$. Because of Eq.~(\ref{lambda<-->x}), the simplicity of the dominant eigenvalue (proved in the first Section of this SM) implies that $y^{(n)}_0(q)$ is a simple root of $Q_n(y,q)$ for $0 \leq q \leq 1$. Therefore
\begin{equation}
  \partial_y Q_n (y^{(n)}_0(q),q) \neq 0 .
  \label{nemelfajulasi_kriterium}
\end{equation}
As a consequence of the implicit function theorem, $y^{(n)}_0(q)$ is locally analytic, implying that the dominant eigenvalue, and therefore the free energy, are analytic functions of $\rho$ within each density regime. When $\delta/d$ crosses an integer value, however, the corresponding value of $n$ changes and the defining polynomial $Q_n$ is replaced by a different one. As a result, the eigenvalue branches generally fail to connect analytically across adjacent density regimes, leading to a nonanalytic dependence of the free energy on $\rho$.

We demonstrate here that the free energy is $(n-1)$-times continuously differentiable at $\rho_n$, while its $n^{\text{th}}$ derivative exhibits a discontinuity. Hence, according to the Ehrenfest classification, $\rho_n$ corresponds to a phase transition of order $n$.

Since the density $\rho_n$ corresponds to the parameter values $(n,q=0)$ and simultaneously to $(n-1,q=1)$, we need to prove that
\begin{equation}
    \left.\frac{d^m y^{(n)}_0}{dq^m}\right|_{q=0}
  = \left.\frac{d^m y^{(n-1)}_0}{dq^m}\right|_{q=1},
    \qquad 0\le m<n,
 \label{m<n_derivaltak}
\end{equation}
while
\begin{equation}
    \left.\frac{d^n y^{(n)}_0}{dq^n}\right|_{q=0}
  \ne
    \left.\frac{d^n y^{(n-1)}_0}{dq^n}\right|_{q=1}.
 \label{n._derivalt}
\end{equation}

Starting from the definition (\ref{Q}), it is straightforward to verify by direct calculation that
\begin{equation}
    Q_n(y,q) = yQ_{n-1}(y,q+1) + (-1)^n\frac{q^n}{n!}.
\label{Q-eq}
\end{equation}
Substituting $q=0$ yields
\begin{equation}
    Q_n(y,0) = yQ_{n-1}(y,1),
\label{folytonos}
\end{equation}
from which it follows immediately that every root of $Q_{n-1}(y,1)$ is also a root of $Q_n(y,0)$. In other words,
\begin{equation}
    y_k^{(n-1)}(1) = y_k^{(n)}(0) , \qquad (k=0,\dots,n-2) .
\end{equation}
Hence, the corresponding branches are continuous across the transition point $(n-1,1)\equiv(n,0)$, therefore, regarded as a function of the density, $\lambda_k$ are continuous at $\rho_n$. In particular, the dominant branch $y_0$ is continuous, and therefore so are $\lambda_0$ and the free energy. Furthermore, since $Q_n(y,0)$ is a polynomial of one degree higher than $Q_{n-1}(y,1)$, it must possess one additional root. It is evident from (\ref{folytonos}) that this extra, newly appearing $n$th root, which is not a root of $Q_{n-1}(y,1)$, is
\begin{equation}
    y_{n-1}^{(n)}(0) = 0.
\end{equation}

Since, by definition, $Q_n(y_k^{(n)}(q),q)=0$, Eq.~(\ref{Q-eq}) implies
\begin{equation}
    Q_{n-1}\bigl(y^{(n)}_k(q),q+1\bigr)
  = \frac{(-1)^{n+1}}{y^{(n)}_k(q)}\frac{q^n}{n!} ,
 \label{Q_Taylor}
\end{equation}
where we have assumed that $y^{(n)}_k(q) \neq 0$, which is necessarily true for the largest solution, $k=0$.

Let us regard $Q_{n-1}(y,q+1)$ as a function of $y$, and expand it in a Taylor series around the point $y_k^{(n-1)}(q+1)$, where by definition this function vanishes. Evaluating this expansion at $y_k^{(n)}(q)$ gives
\begin{align}
 Q_{n-1}\bigl(y^{(n)}_k(q),q+1\bigr)
  = A \bigl(y^{(n)}_k(q)-y^{(n-1)}_k(q+1)\bigr)
    +\mathcal{O}\!\left(
    \bigl(y^{(n)}_k(q)-y^{(n-1)}_k(q+1)\bigr)^2
    \right),
 \label{Taylor_Q}
\end{align}
where, for convenience, we have introduced the notation
\begin{equation}
    A = \partial_yQ_{n-1} \bigl(y^{(n-1)}_k(q+1),q+1\bigr).
\end{equation}
In the following we suppose that $A\neq0$, which is necessarily true for the largest solution, see Eq.~(\ref{nemelfajulasi_kriterium}).
Comparing Eqs.~(\ref{Q_Taylor}) and (\ref{Taylor_Q}), we first obtain that $y^{(n)}_k(q)-y^{(n-1)}_k(q+1)=\mathcal{O}(q^n)$. Consequently, the quadratic remainder term is $\mathcal{O}(q^{2n})=o(q^n)$, which yields 
\begin{align}
    y^{(n)}_k(q)-y^{(n-1)}_k(q+1)
    =
    \frac{(-1)^{n+1}}
         {n!\,A\,y^{(n)}_k(q)}
    q^n
    +o(q^n) .
 \label{y_Taylor}
\end{align}

Equation (\ref{y_Taylor}) shows, on the one hand, that the Taylor expansions of the two functions about $q=0$ coincide up to order $n-1$. Consequently, Eq.~(\ref{m<n_derivaltak}) holds for every $m<n$.

On the other hand, the difference between the $n^{\text{th}}$ derivatives does not vanish, since the coefficient of $q^n$ on the right-hand side of Eq.~(\ref{y_Taylor}) is nonzero. Thus, we can establish a stronger result than the statement (\ref{n._derivalt}). Taking $n$ derivatives of Eq.~(\ref{y_Taylor}) and evaluating the result at $q=0$, we obtain the magnitude of the discontinuity:
\begin{align}
\left.\frac{d^n y^{(n)}_0}{dq^n}\right|_{q=0}
-
\left.\frac{d^n y^{(n-1)}_0}{dq^n}\right|_{q=1}
=
  \frac{(-1)^{n+1}}
     {\partial_yQ_{n-1} \bigl(y^{(n-1)}_0(1),1\bigr)\,y^{(n)}_0(0)}
\neq 0.
\end{align}


\section{Grand canonical solution using density functional theory for regimes $n=1$ and $n=2$}

\subsection{Geometric setup and Percus' functional}

As described in the main text,
we consider hard rods of diameter $\sigma$ on a line, subject to a periodic
external potential of period $\ell$ consisting of square wells of width
$\delta<\sigma$ (so that at most one rod occupies a well). The well
occupying the reference cell is centered at the origin, with
$\rho(x)=0$ for $|x|>\delta/2$.

We treat the grand canonical problem with DFT where the equilibrium profile is found by minimizing the
functional for the grand free energy,
\begin{equation}
  \Omega[\rho] = F_\rmid[\rho] + F_\rmex[\rho] + \int dx (\vext(x) -\mu) \rho(x)\;,
\end{equation}
with
\begin{equation}
 \beta F_\rmid[\rho] = \int dx \rho(x) \left( \ln[\rho(x) \Lambda] -1 \right) \;.
\end{equation}
The exact excess free energy of the 1D hard-rod fluid can be written 
in an asymmetric, one-sided form,
\begin{equation}
  \beta F_\rmex[\rho] \;=\; \int dx \, \Big[-\rho(x)\,\ln\!\big(1-\eta(x)\big)\Big],
  \qquad
  \eta(x) \;=\; \int_x^{x+\sigma}\rho(y)\,dy .
  \label{eq:Fex_asym}
\end{equation}

We define the activity by $z=\exp(\beta\mu)$. The limit $z\to\infty$ drives the wells to maximal occupation ($N_{\rm cell}=1$) and recovers 
the model studied in the main text. In the following, $\beta=1$ since only hard interactions are involved, and $\Lambda=1$. 

\textit{Uncorrelated regime:} This is the regime $n=1$ from the main text which holds for $\ell \ge \sigma +\delta$.
Here, from minimizing $\Omega$,  one finds a flat density $\rho_0=N/\delta$ with $N\equiv N_{\rm cell}=z\delta/(1+z\delta)$ and the free energies
\begin{align}
  F_\rmid &= N\left(\ln\frac{N}{\delta}-1\right), \\
  F_\rmex &= N + (1-N)\ln(1-N).
  \label{eq:Fex_isolated}
\end{align}
$F_\rmex$ is the free energy of a 0D cavity which can hold one particle at most and therefore depends only on $N$. In the limit
$z\to\infty$ we find $N=1$ and $F=F_\rmid+F_\rmex=-\ln\delta$, as in the main text.

The regime of interest is the \emph{correlated} one (regime $n=2$ in the notation of the main text), where the gap between
neighboring wells is smaller than the rod diameter, so that a rod near the
edge of one well can hard-core exclude a rod near the edge of the
neighboring well:
\begin{equation}
  \ell-\sigma \;\in\; \left(\tfrac{\delta}{2},\,\delta\right).
\end{equation}
It is convenient to define
\begin{equation}
  w \;=\; \delta-(\ell-\sigma), \qquad w\in\left(0,\tfrac{\delta}{2}\right),
  \qquad
  \Delta_1 \;=\; \tfrac{\delta}{2}-w,
\end{equation}
so that $\Delta_1$ is the boundary between a flat central region and two edge
regions of the well, and $w\to0$ recovers the isolated-well limit
$\ell\to\delta+\sigma$.

\subsection{Ansatz for $\rho(x)$ for regime $n=2$}

Motivated by the exact $\mu\to\infty$ (one rod per well) transfer-matrix
solution, we consider the piecewise flat + linear ansatz, for $x>0$
(extended to $x<0$ by $\rho(-x)=\rho(x)$):
\begin{equation}
  \rho(x) =
  \begin{cases}
	  A, & 0\le x\le \Delta_1 \qquad(\text{flat region})\;, \\[4pt]
    A+B\,(x-\Delta_1), & \Delta_1\le x\le \delta/2  \qquad(\text{edge region})\; .
  \end{cases}
  \label{eq:ansatz}
\end{equation}

\subsection{Euler--Lagrange equation and its differential form}

The exact (undifferentiated) Euler--Lagrange equation following from
\eqref{eq:Fex_asym} is
\begin{equation}
  \ln\rho(x) - \mu - \ln\big(1-\eta(x)\big)
  + \int_{x-\sigma}^{x} \frac{\rho(y)}{1-\eta(y)}\,dy \;=\; 0 .
  \label{eq:ELE}
\end{equation}
Differentiating with respect to $x$ (using Leibniz's rule on $\eta(x)$ and on
the integral's limits) gives the compact local relation
\begin{equation}
  \frac{\rho'(x)}{\rho(x)} \;=\; \frac{\rho(x-\sigma)}{1-\eta(x-\sigma)}
  \;-\; \frac{\rho(x+\sigma)}{1-\eta(x)} .
  \label{eq:ELE_diff}
\end{equation}

\paragraph*{Consistency in the flat region.}
For $0<x<\Delta_1$: the left-hand side vanishes trivially, since $\rho(x)=A$ is
constant. On the right-hand side, both $\rho(x-\sigma)$ and $\rho(x+\sigma)$
vanish: $x-\sigma$ lies strictly inside the empty gap to the left (it has
passed both the own well's left edge and stops short of the left
neighbor), and $x+\sigma < \ell-\delta/2$ for $x<\Delta_1$ (using
$\Delta_1+\sigma = \ell-\delta/2$), so $x+\sigma$ has not yet reached the right
neighbor either, while it has already passed the own well's right edge
since $\delta<\sigma$. Hence RHS $=0$ as well, and \eqref{eq:ELE_diff} is
satisfied identically throughout the flat region --- a necessary but not
sufficient check (it fixes no relation between $A$ and $B$, since a
differentiated equation cannot see the integration constant).

\subsection{Solution in the edge region}

For $\Delta_1<x<\delta/2$: $\rho(x-\sigma)=0$ still holds (same argument as above),
while $\rho(x+\sigma)=A+B(\delta/2-x)$ now picks up the neighboring well's
edge. Equation~\eqref{eq:ELE_diff} becomes, with $v=x-\Delta_1\in(0,w)$,
\begin{equation}
  \frac{B}{A+Bv} \;=\; -\,\frac{A+B(w-v)}{1-\eta(x)},
  \qquad
  \eta(x) = Aw + B\left(\frac{w^2}{2}+wv-v^2\right).
\end{equation}
Multiplying through by $\rho(x)(1-\eta(x))$ and expanding, the terms linear
and quadratic in $v$ cancel \emph{identically}, 
leaving a single $v$-independent condition:
\begin{equation}
  \boxed{\;A^2 + B - \frac{B^2w^2}{2} \;=\; 0\;}
  \label{eq:edge_condition}
\end{equation}

\subsection{Solution in the flat region and the second condition}

Returning to the full, undifferentiated Euler-Lagrange equation~\eqref{eq:ELE} for $x$ in
the flat region: the backward integral splits at $y=-\Delta_1$ into an
$x$-independent piece and an $x$-dependent piece,
\begin{equation}
  \int_{x-\sigma}^{x} \frac{\rho(y)}{1-\eta(y)}\,dy
  \;=\; \underbrace{\int_{-\delta/2}^{-\Delta_1}\frac{\rho(y)}{1-\eta(y)}\,dy}_{I_{\rm left}}
  \;+\; \int_{-\Delta_1}^{x}\frac{\rho(y)}{1-\eta(y)}\,dy ,
\end{equation}
using that $\rho(y)=0$ for $y<-\delta/2$, together with $x-\sigma<-\delta/2$
always holding here. In the flat sub-region, $1-\eta(y)=1-A(\delta/2-y)-Bw^2/2$
has derivative $A$, exactly matching the numerator $\rho(y)=A$, so the second
piece is a pure log-derivative:
\begin{equation}
  \int_{-\Delta_1}^{x}\frac{A}{1-\eta(y)}\,dy \;=\; \ln\big(1-\eta(x)\big)-\ln\big(1-\eta(-\Delta_1)\big).
\end{equation}
The resulting $+\ln(1-\eta(x))$ exactly cancels the equation's own
$-\ln(1-\eta(x))$ term. A second, analogous simplification occurs in
$I_{\rm left}$: throughout the left-edge sub-region, $\rho(y+\sigma)=0$, so
$\eta'(y)=\rho(y+\sigma)-\rho(y)=-\rho(y)$ and again
$\rho(y)/(1-\eta(y))=\frac{d}{dy}\ln(1-\eta(y))$ identically, giving
$I_{\rm left}=\ln(1-\eta(-\Delta_1))-\ln(1-\eta(-\delta/2))$. The $\ln(1-\eta(-\Delta_1))$
terms cancel a second time, leaving the remarkably simple, exactly
$x$-independent condition
\begin{equation}
  \ln A \;=\; \mu + \ln\big(1-\eta(-\delta/2)\big).
\end{equation}
One shows $\eta(-\delta/2)=N_{\rm cell}\equiv A\delta+Bw^2$ (the window
starting at the well's own edge and reaching forward by $\sigma>\delta$
captures exactly the whole well and nothing else). 
\begin{equation}
  \boxed{\;A \;=\; z\big(1-N_{\rm cell}\big) \;=\; z\big(1-A\delta-Bw^2\big)\;}
  \label{eq:flat_condition}
\end{equation}

\subsection{Full solution for $A$ and $B$}

Solving \eqref{eq:flat_condition} for $A$ in terms of $B$ and substituting
into the edge condition \eqref{eq:edge_condition} gives a quadratic equation
for $B$; of its two roots, only one gives physically sensible ($A>0$,
modest $B$) values. Defining
\begin{equation}
  D \;=\; (1+\delta z)^2 - 2w^2z^2 ,
\end{equation}
the physical solution simplifies remarkably to
\begin{equation}
  \boxed{\;A \;=\; \frac{z}{\sqrt{D}}\;},
  \qquad
  \boxed{\;B \;=\; \frac{1}{w^2}\left(1-\frac{1+\delta z}{\sqrt{D}}\right)\;},
  \qquad
  \boxed{\;N_{\rm cell} \;=\; 1-\frac{1}{\sqrt{D}}\;}.
  \label{eq:AB_solution}
\end{equation}
As $w\to0$, $D\to(1+\delta z)^2$ and $N_{\rm cell}\to \delta z/(1+\delta z)$,
recovering the isolated-well result exactly. Since $\sqrt{D}<1+\delta z$ for
any $w\neq0$, correlation always \emph{reduces} $N_{\rm cell}$ relative to
the isolated well.

We verified that $z\to\infty$ recovers the TOM solution in the main text for the regime $n=2$.

\subsection{Ideal and excess free energy}



Integrating the ansatz \eqref{eq:ansatz} directly gives the ideal part in
closed form:
\begin{equation}
  F_\rmid = -A^2\ln A + A^2\ln(A+Bw) + AB(\delta-2w)(\ln A - 1)
  - \frac{w\big(6A+3Bw-2(2A+Bw)\ln(A+Bw)\big)}{2}.
  \label{eq:Fid_correlated}
\end{equation}
The excess part requires all three sub-regions (flat region and left/right edge region) separately, since $\eta(x)$
(unlike $\rho(x)$) is \emph{not} mirror-symmetric:

\begin{equation}
  F_\rmex = F_{\rmex,\rm flat} + F_{\rmex,\rm right} + F_{\rmex,\rm left}, \qquad
  F_\rmex = \int_{-\delta/2}^{\delta/2}\big[-\rho(x)\ln(1-\eta(x))\big]\,dx .
\end{equation}

\paragraph{A key structural simplification.} In the flat and left-edge
regions, $\rho(x+\sigma)=0$ throughout (established before), so
$\eta'(x)=\rho(x+\sigma)-\rho(x)=-\rho(x)$, i.e.\
$\rho(x)=\tfrac{d}{dx}[1-\eta(x)]$ \emph{exactly}. Substituting $V=1-\eta(x)$,
the integral collapses to the elementary form $\int\ln(V)\,dV=V\ln V-V$.
The right-edge region lacks this property ($\rho(x+\sigma)\neq0$ there,
since it picks up the neighboring well), so it still requires the fuller
treatment below. Define
\begin{equation}
  \eta_{\rm edge} \equiv \eta(x_c) = \eta(\delta/2) = Aw+\tfrac{Bw^2}{2},
  \qquad
  \eta_c \equiv \eta(-x_c) = A(\delta-w)+\tfrac{Bw^2}{2},
\end{equation}
(the first equality in each definition reflects boundary values of $\eta$
shared between adjacent regions; $\eta_{\rm edge}$ additionally takes the
\emph{same} value at both ends of the right-edge region, a non-obvious
coincidence of this geometry). Then, exactly:
\begin{align}
  F_{\rmex,\rm flat} &= (1-\eta_c)\ln(1-\eta_c) - (1-\eta_{\rm edge})\ln(1-\eta_{\rm edge}) + \eta_c - \eta_{\rm edge}, \\
  F_{\rmex,\rm left} &= (1-N_{\rm cell})\ln(1-N_{\rm cell}) - (1-\eta_c)\ln(1-\eta_c) + N_{\rm cell} - \eta_c .
\end{align}
The right-edge piece was found 
via integration by parts, $\int\rho\,dv=V(v)=Av+Bv^2/2$, reducing it to a
polynomial part plus the standard linear-over-quadratic integral. 
After simplification: 
\begin{eqnarray}
	F_{\rmex,\rm right} &=& -\eta_{\rm edge}\ln(1-\eta_{\rm edge}) + 2\eta_{\rm edge}
  + \frac{(2A+Bw)\sqrt{\mathrm{disc}}}{4B}\,
  \ln\left|\frac{2Bw-\sqrt{\mathrm{disc}}}{2Bw+\sqrt{\mathrm{disc}}}\right| ;, \\
	 & & \mathrm{disc}=4B(4Aw+3Bw^2-4)
\end{eqnarray}

\paragraph{The total.} Adding all three, the $\eta_c$ terms cancel
completely (a telescoping between the flat and left-edge pieces), and the
two remaining $\ln(1-\eta_{\rm edge})$ contributions (from the flat and
right-edge pieces) combine into one:
\begin{equation}
  \boxed{\,F_\rmex = (1-N_{\rm cell})\ln(1-N_{\rm cell}) + N_{\rm cell}
  - \ln(1-\eta_{\rm edge}) + \eta_{\rm edge}
  + \frac{(2A+Bw)\sqrt{\mathrm{disc}}}{4B}\,
  \ln\left|\frac{2Bw-\sqrt{\mathrm{disc}}}{2Bw+\sqrt{\mathrm{disc}}}\right|\,}
  \label{eq:Fex_correlated_total}
\end{equation}
The first two terms are exactly the isolated-well $F_\rmex$
\eqref{eq:Fex_isolated} evaluated at $N_{\rm cell}$; the entire correlation
correction is contained in the remaining three terms. All individual
pieces and the total were verified against direct numerical integration of
the ansatz.


\subsection{Numerical example: $\delta=0.5$, $\beta\mu=1.0$, $\sigma=1$}

\begin{figure}[htbp]
  \centering
  \includegraphics[width=0.95\textwidth]{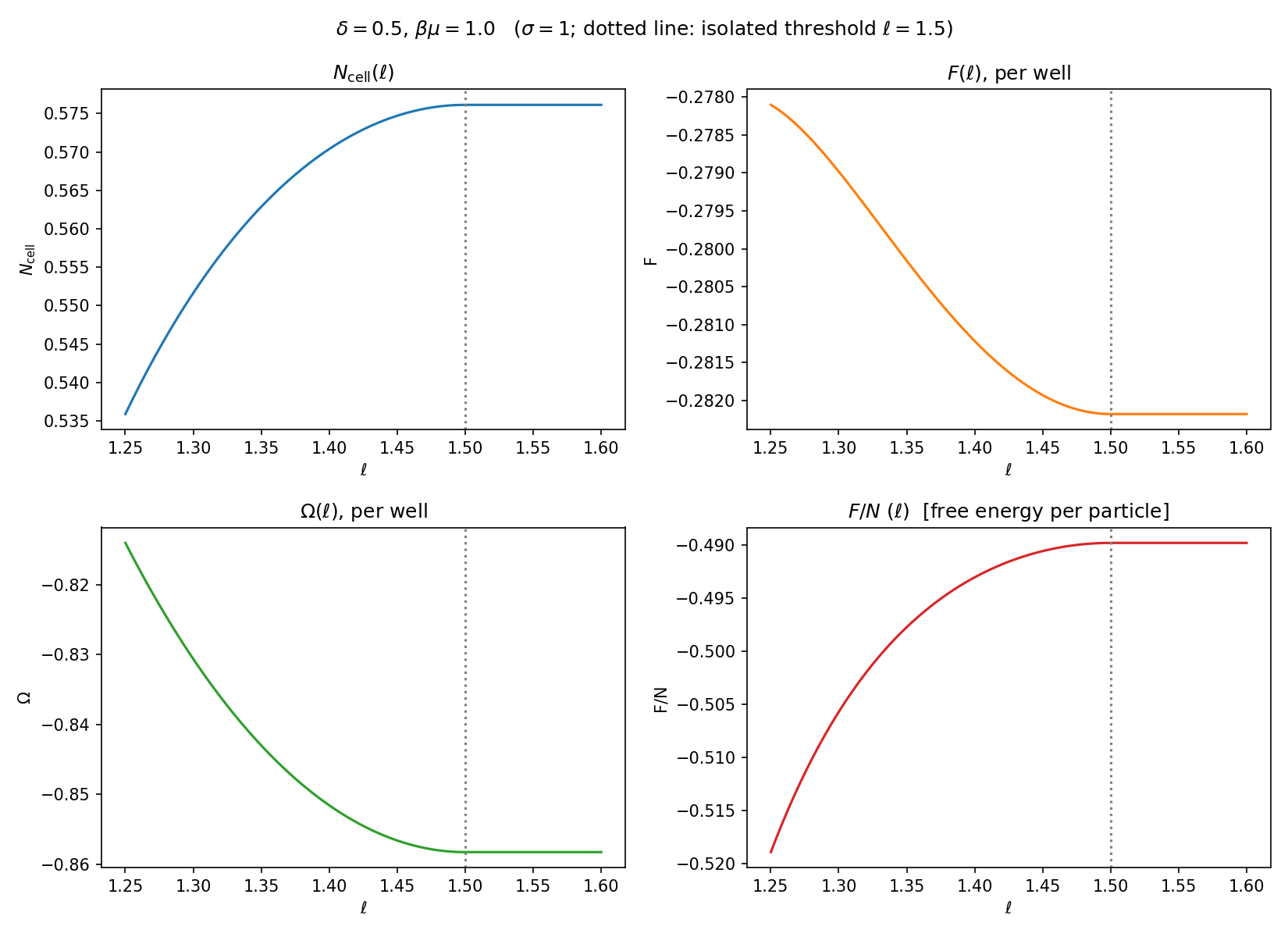}
  \caption{$N_{\rm cell}(\ell)$, total $F(\ell)$, total $\Omega(\ell)$, and $F/N(\ell)$
  per well, for $\delta=0.5$, $\beta\mu=1.0$. The dotted line marks the
  isolated-well threshold $\ell=1.5$.}
  \label{fig:overview}
\end{figure}

Figure~\ref{fig:overview} shows $N_{\rm cell}(\ell)$, $F(\ell)$, $\Omega(\ell)$, and
$F/N(\ell)$ (free energy per particle, $a\equiv F/N$) across the correlated
regime $\ell\in(1.25,1.5)$ and into the isolated plateau $\ell\in[1.5,1.6]$,
computed from the closed-form solution for the density, \eqref{eq:AB_solution} together with
direct numerical quadrature of $F_\rmid,F_\rmex$.
All four quantities transition smoothly, by eye, into the flat
isolated-well value at $\ell=1.5$.

\begin{figure}[htbp]
  \centering
  \includegraphics[width=\textwidth]{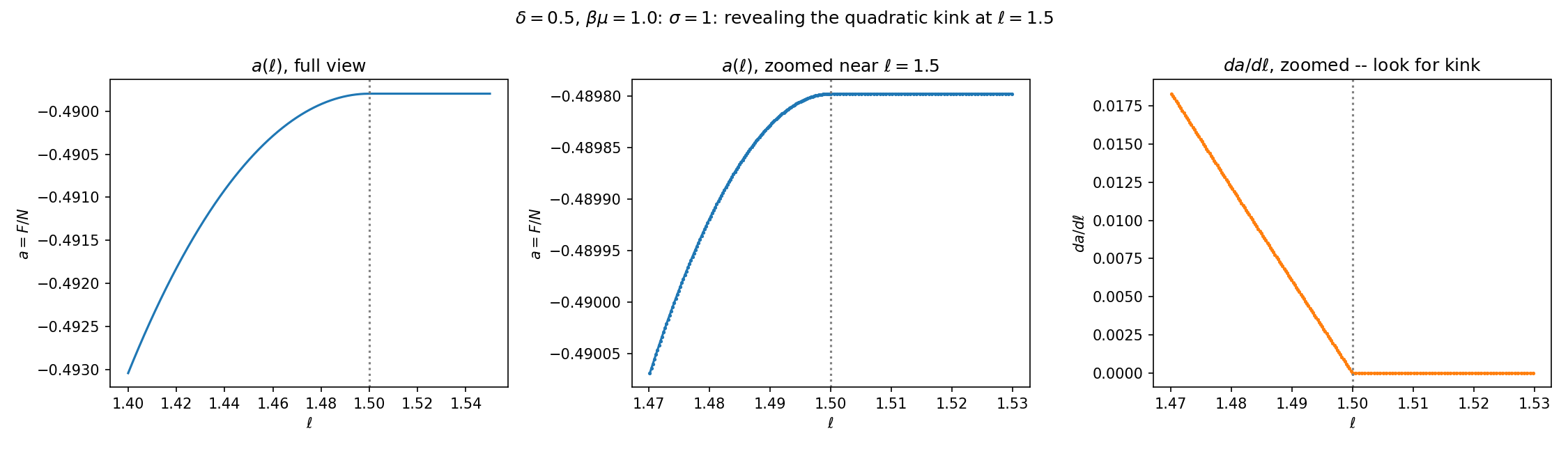}
  \caption{$a=F/N$ near the threshold $\ell=1.5$: full view, zoomed view, and
  $da/d\ell$ zoomed --- the linear approach to zero from below, meeting a flat
  zero above, reveals the quadratic kink.}
  \label{fig:kink}
\end{figure}

Zooming in near $\ell=1.5$ and examining the first derivative reveals a
non-analyticity invisible in Fig.~\ref{fig:overview}: $a(\ell)=F/N(\ell)$ is
continuous with a continuous first derivative, but $da/d\ell$ approaches zero
\emph{linearly} from the correlated side and meets a flat zero on the
isolated side --- a kink signaling a jump in the second derivative
(Fig.~\ref{fig:kink}). This matches the exact transfer-matrix result derived
earlier for the $\mu\to\infty$ limit, $a-a_c\propto(\delta+\sigma-\ell)^2$ near
threshold, now confirmed to persist at finite $\mu$.

\begin{figure}[htbp]
  \centering
  \includegraphics[width=\textwidth]{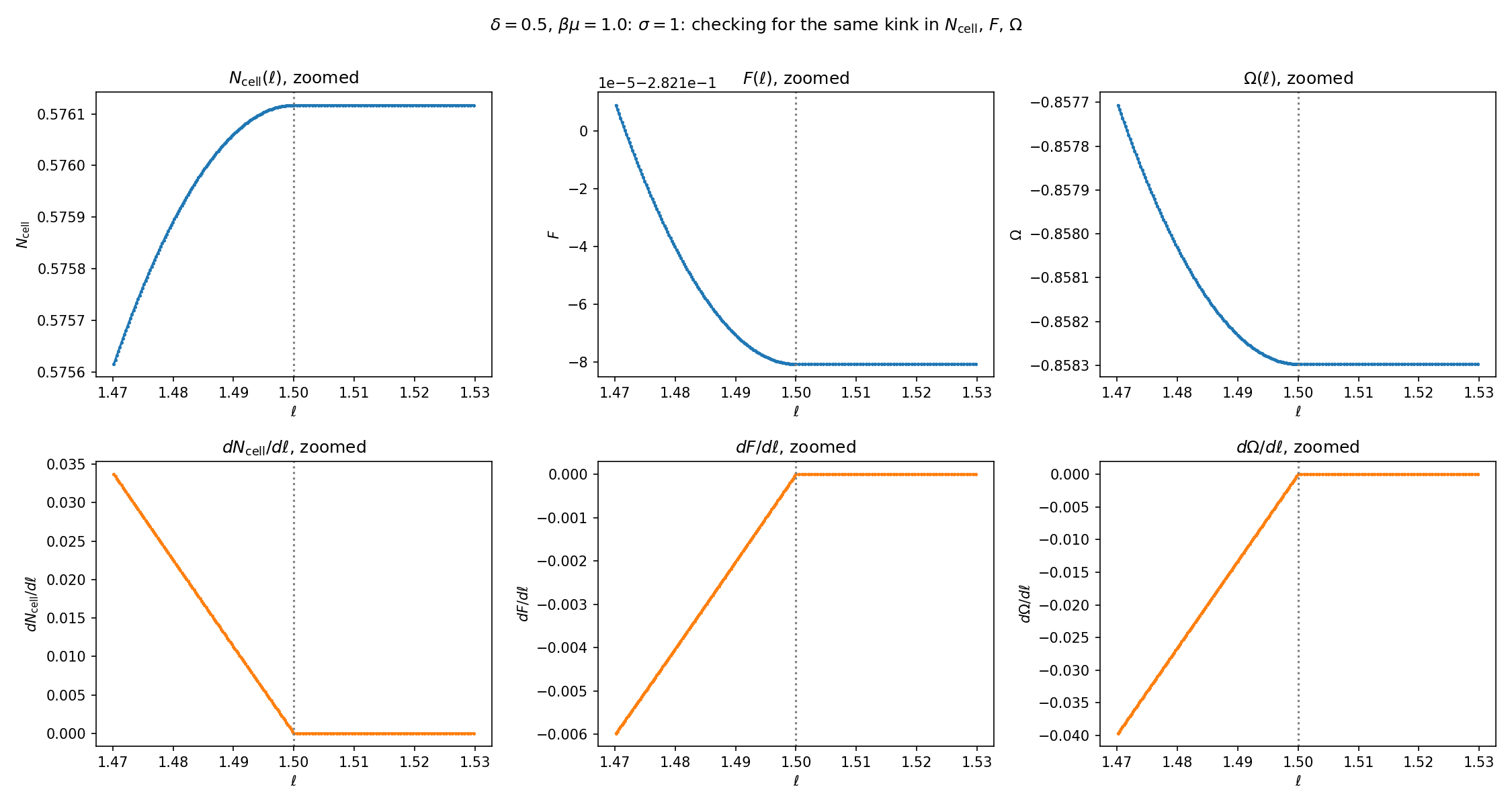}
  \caption{$N_{\rm cell}(l)$, $F(l)$, $\Omega(l)$ (top row) and their
  derivatives $dN/dl$, $dF/dl$, $d\Omega/dl$ (bottom row), zoomed near
  $\ell=1.5$. All three show the same kink signature as $a=F/N$.}
  \label{fig:kink_all}
\end{figure}

The same kink structure is not special to the ratio $a=F/N$: it is
inherited from $N_{\rm cell}(\ell)$, $F(\ell)$, and $\Omega(\ell)$ individually, each
of which shows the identical linear-derivative-to-zero signature at
$\ell=1.5$ (Fig.~\ref{fig:kink_all}).

\bibliography{references}
